\pdfoutput=1

\documentclass[10pt,a4paper]{bmcart-tuned}

\usepackage[bmc]{bmcart-biblio}

\usepackage[utf8]{inputenc} 
\usepackage[english]{babel} 

\usepackage{blindtext}

\usepackage[scaled=.90]{helvet}

\usepackage[sc]{mathpazo} \usepackage[T1]{fontenc} 

\usepackage[activate={true,nocompatibility},final,tracking=true,kerning=true,spacing=true,factor=1100,stretch=10,shrink=10]{microtype}

\usepackage{enumitem} \setlist[itemize]{noitemsep} 
\usepackage{epigraph}

\usepackage{lettrine}

\usepackage{abstract}   

\usepackage{titlesec} \renewcommand\thesection{\Roman{section}}  \titleformat{\section}[block]{\large\centering\bfseries}{\thesection.}{1em}{\vspace{0.3em}} \titleformat{\subsection}[block]{\normalfont\large\bfseries}{}{0.0em}{\vspace{0.15em}} 
\usepackage{adjustbox}

\usepackage{booktabs}

\usepackage{array,tabularx} 
\usepackage[flushleft]{threeparttable}

\makeatletter
\appto\@floatboxreset{  \ifx\@captype\andy@table
    \normalfont
      \fi
}
\def\andy@table{table}
\makeatother

\usepackage{calc}

\usepackage{footmisc}

\usepackage{stix}

\usepackage[hyphens,spaces,obeyspaces]{url}
\RequirePackage{etoolbox} \RequirePackage{hyperref} 
\newrobustcmd{\wptitle}{How do Software Ecosystems Co-Evolve?}
\pdfstringdef{\wptitlePDF}{How do Software Ecosystems Co-Evolve?}

\newrobustcmd{\wpsubtitle}{A view from OpenStack and beyond}
\pdfstringdef{\wpsubtitlePDF}{A view from OpenStack and beyond}

\pdfstringdef{\wpsubject}{Eighth International Conference on Software Business, Essen 2017}
\pdfstringdef{\wpsubjectPDF}{Eighth International Conference on Software Business, Essen 2017}

\newrobustcmd{\wpasin}{As accepted for presentation at the 8th International Conference on Software Business (ICSOB 2017), held in Essen, Germany,
June, 12-13, 2017. The official conference proceedings are  available at \url{www.springerlink.com}.}

\newrobustcmd{\wprunningauthors}{Apolinário Teixeira and Hyrynsalmi}
\pdfstringdef{\wprunningauthorsPDF}{Apolinário Teixeira and Hyrynsalmi}

\pdfstringdef{\wpkeywordslist}{alliances, business ecosystem, software ecosystem, evolution, coopetition, open-coopetition, open-source, OpenStack.}

\newrobustcmd{\wppaurl}{\url{http://www.jteixeira.eu/pub/wp/ICSOB2017SEESNA.pdf}}

\RequirePackage{etoolbox}

\newtoggle{bibtex_natbib}
\settoggle{bibtex_natbib}{false}

\newtoggle{biber_biblatex}
\settoggle{biber_biblatex}{true}

\iftoggle{bibtex_natbib}{
  \typeout{On bibtex_natbib mode for bibliographies}
}{}

\iftoggle{biber_biblatex}{
  \typeout{On biber_biblatex mode for bibliographies}
}{}

\iftoggle{bibtex_natbib}{
  \iftoggle{biber_biblatex}{
    \typeout{ERROR: Chose among packages and pre-processors of bibliographies}
    \typeout{Either bibtex_natbib or biber_biblatex}
    \stop}{} 
}{}

\iftoggle{bibtex_natbib}{
}{ \iftoggle{biber_biblatex}{}
  {
    \typeout{ERROR: Chose one package and pre-processor of bibliographies}
    \typeout{Either bibtex_natbib or biber_biblatex}
    \stop
  }
}

\startlocaldefs

\iftoggle{biber_biblatex}{

      \makeatletter
  \let\c@author\relax
  \makeatother

  \usepackage[style=numeric,backend=biber,sorting=none]{biblatex}
        \addbibresource{./references.bib}
    \addbibresource{./ecosystem.bib}
    \addbibresource{./hyrynsalmi.bib}
    \addbibresource{./bibliographicreferencesforwritingacademicpapers/JoseTeixeiraPub.bib}
  }{}

\iftoggle{bibtex_natbib}{
    \RequirePackage[sort&compress,numbers]{natbib}
  }

\usepackage[autostyle]{csquotes}

\usepackage{theorem}

\usepackage{pdfpages}

\usepackage{cleveref}

 \usepackage[]{hyperref}
  \hypersetup{
    pdfinfo={  
    Title={\wptitlePDF\ \wpsubtitlePDF},  
    Subject={\wpsubject},  
    Author  = {\wprunningauthorsPDF}, 
    Keywords  ={\wpkeywordslist},   
  }
}

\endlocaldefs

\begin{document}
 \pagenumbering{roman}
 \pagestyle{empty}
 \includepdf[pages={1-},scale=0.99, 
 addtotoc={1,section,1,Cover,p1,2,section,1,Award,p2,3,section,1,Copyright,p3,4,section,1,Funding,p4,5,section,1,Formatting,p5}]{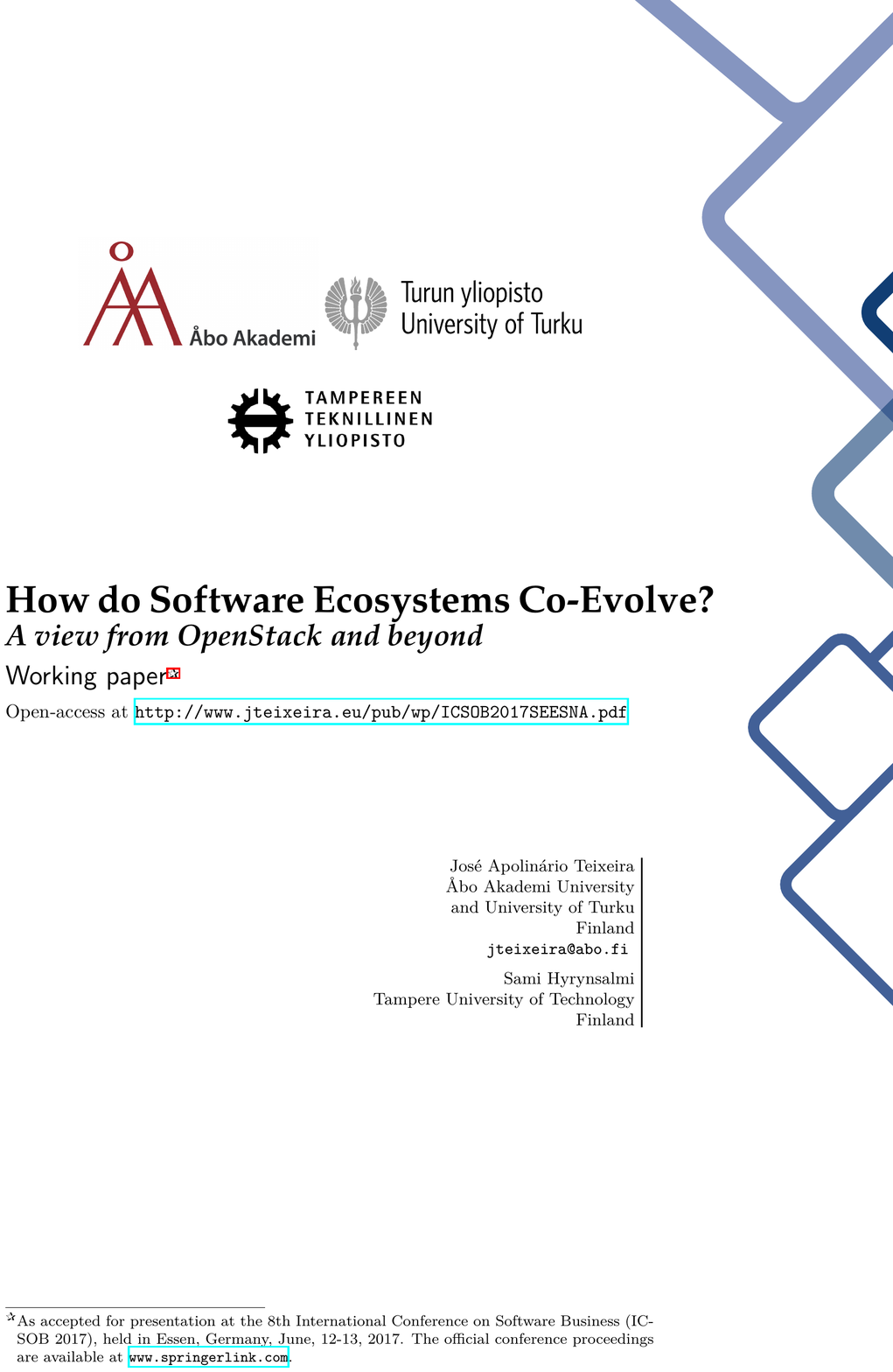}

\cleardoublepage

\setcounter{section}{0}
\renewcommand*{\theHsection}{chY.\the\value{section}}

\pagenumbering{arabic}
\setcounter{page}{1}
\pagestyle{bmcheadings}

\begin{frontmatter}

\begin{fmbox}
\dochead{Research working paper}

\title{\wptitle}
\subtitle{\wpsubtitle{\Large\thanks{\footnotesize \wpasin}}}

\author{\inits{JAT}\fnm{Jose} \snm{Apolinário Teixeira}\textsuperscript{\wpAuthCorrespondant,\wpAuthSymbFirst,\wpAuthSymbSecond}}
\author{\inits{SH}\fnm{Sami} \snm{Hyrynsalmi}\textsuperscript{\wpAuthSymbSecond}}

\newcommand{\wpCorrespondance}{{\color{joseViolet2016}{$\star$}} For correspondence (academic issues only): {\color{blue}jose.teixeira@utu.fi}}

\newcommand{\wpAffiliations}{\wpAuthSymbFirst  University of Turku, Finland  
\newline \wpAuthSymbSecond Åbo Akademi University, Finland 
\newline \wpAuthSymbThird Tampere University of Technology, Finland}

\setattribute{myauthorinfo}       {text} { \vspace{0.5em} \wpCorrespondance \vspace{0.3em} \newline \wpAffiliations \endgraf}
\setattribute{copyright}        {text} { \wprunningauthors  \hfill } 
\setattribute{runninghead}      {text} { \wprunningauthors  \hfill Page\ \thepage\  of \pageref{LastPage}}

\begin{artnotes}
\end{artnotes}
\end{fmbox}

\begin{abstractbox}

\begin{abstract}

Much research that analyzes the evolution of a software ecosystem is confined to its own boundaries.  Evidence shows, however, that software ecosystems co-evolve independently with other software ecosystems. In other words,  understanding the evolution of a software ecosystem requires an especially astute awareness of its competitive landscape and much consideration for other software ecosystems in related markets. A software ecosystem does not evolve in insulation but with other software ecosystems.
In this research, we analyzed the OpenStack software ecosystem with a focal perspective that attempted to understand its evolution as a function of other software ecosystems. We attempted to understand and explain the evolution of OpenStack in relation to other software ecosystems in the cloud computing market. Our findings add to theoretical knowledge in software ecosystems by identifying and discussing seven different mechanisms by which software ecosystems mutually influence each other: sedimentation and embeddedness of business relationships, strategic management of the portfolio of business relationships, firms values and reputation as a partner, core technological architecture, design of the APIs, competitive replication of functionality and multi-homing.
Research addressing the evolution of software ecosystem should, therefore, acknowledge that software ecosystems entangle with other software ecosystems in multiple ways, even with competing ones. A rigorous analysis of the evolution of a software ecosystem should not be solely confined to its inner boundaries.

\end{abstract}


\end{abstractbox}

\end{frontmatter}

\section{Introduction}

\lettrine[]{In}{the}
so-called `Information Age' companies and organizations do not live in isolation; instead, business activities of modern companies are highly interwoven with other companies. Furthermore, the fate of a company nowadays depends on its connections and environment where they are working --- not anymore solely on the company itself. From these observations, James F.\ Moore \cite{Moore1993} built his theory and concept of `business ecosystem'. According to Moore \cite{Moore1996}, a business ecosystem consists of a set of companies working on a shared innovation. The companies work together, cooperatively and competitively, for creating value for customers; the ecosystem advances as companies and the innovation co-evolve together.

Since Moore's seminal article, a plethora of different kinds of artificial ecosystems has been defined and used \cite{Suominen2016a}. One of the most important is `software ecosystem' as software is pervasive and ubiquitous by its nature --- there hardly is any  industrial domain where software would not be a part of. That is, software is available nowadays everywhere and it is rarely built on isolation. Software ecosystems have become also an important research field and there are hundreds of studies addressing different kinds of software ecosystems \cite[c.f.][]{Manikas2013a, Manikas2016a}.

As proposed by Jansen et al.~\cite{Jansen2009a}, a software ecosystem consists of ``\emph{the set of businesses functioning as a unit and interacting with a shared market for software and services, together with the relationships among them. These relationships are frequently underpinned by a common technological platform or market and operate through the exchange of information, resources and artefacts.}'' From the definition, it seems clear that there are a relationship between software ecosystem and business ecosystem conceptualizations. However, there is one major caveat: Whereas Moore's view on business ecosystems focused on co-evolution, the definition of software ecosystem does not cover this aspect. 

Since the ecosystem concept has been accepted as a perspective for business development, management and governance, it is also necessary to discuss the interactions between ecosystems. Whether and how ecosystems influence each other and to what degree, are intriguing questions. That is, it is essential for both scholars and practitioners to analyze the competitive landscape, in order to better understand software ecosystem evolution. This study focuses on the relatively uncovered area in the field of software ecosystems:  the \emph{co-evolution} of them. Specifically, we focus on how ecosystems influence each other evolution (i.e.\ co-evolution of ecosystems). The starting research hypothesis is that an ecosystem does not evolve in isolation; instead, the ecosystems are interwoven with each other (e.g., characteristics of one can affect the other or change in one can also affect the other). That is, this study seeks to answer a question
\begin{description}
\item[RQ] \emph{How do  software ecosystems co-evolve?}
\end{description}

To answer the presented research question, we performed a case study by taking the OpenStack software ecosystem as our focal unit of analysis. By analyzing OpenStack in relation to other software ecosystems in the industry (e.g., CloudStack) we identified and explored seven different ways in which software ecosystems are interwoven with other  software ecosystems: 
\begin{enumerate}
\item[ $\alpha$)] \emph{Sedimentation and embeddedness of business relationships},
\item[ $\beta$)] \emph{Strategic management of the portfolio of business relationships},
\item[ $\gamma$)] \emph{Firm's values and reputation as a partner}, 
\item[ $\delta$)] \emph{Core technological architecture}, 
\item[ $\epsilon$)] \emph{Design of external APIs},
\item[ $\zeta$)] \emph{Technological replication of new functionalities}, and
\item[ $\eta$)] \emph{Complementors' multi-homing}.
\end{enumerate}

The remaining of this paper is structured as follow. The next section will briefly present the related work whereas Section \ref{sec:empiricalbc} views on the empirical background of our study subject. Section \ref{sec:method} presents the research approach used, Section \ref{sec:results} results and Section \ref{sec:disc} their implications and limitations. The final section concludes the study.

\section{Related literature}
\label{sec:theory}
Moore \cite{Moore1993}, in his seminal essay on the new ecology of competition, defined that business ecosystems evolve through distinct phases. He identified and named four stages which are:
\begin{enumerate}
\item \emph{Birth} where companies define value propositions of a seed innovation.
\item \emph{Expansion} where the ecosystem seeks to expand to new territories. 
\item \emph{Leadership} where participating companies start to struggle for a leadership. As an example, Moore used Microsoft's and Intel campaign against IBM during the ``clone wars'' of personal computers. 
\item \emph{Self-Renewal} or \emph{Death} where an ecosystem faces an external threat and it is forced to either renew itself or cease to exist. 
\end{enumerate}

As software ecosystem share distinct similarity with the older ecosystem concept, it is surprisingly how little have been written on the evolution of different kinds of ecosystems. However, there are a few prior studies existing. Similarly, as have entire software ecosystem literature diverged into communities \cite{Suominen2016a}, also existing studies can be categorized into two groups with a remarkable different basis.

In the first group, there are studies addressing software ecosystem as a business network and previous studies have addressed how the relationships between the firms have developed. For example, Basole \cite{basole_visualization_2009} studied the convergence of entire mobile ecosystem---including software and hardware vendors as well as network operators---however, it focused on the interfirm relationships and visualization of cooperative networks. Basole and Karla \cite{Basole2011} studied the evolution of mobile platform ecosystems. However, also their focus was on the visualization and on the interfirm relationships inside an ecosystem. Hanssen \cite{Hanssen2012b} followed a transformation of a product line organization to an emerging software ecosystem and focused on why and how the transformation was done. 

In the second group, there are studies addressing software ecosystems as a collection of interdependent projects and these studies on the evolution of software ecosystem's codebase over time \cite[e.g.][]{German2013, Bavota2013}. Already 2007, Yu and Bush \cite{Yu2007} noted that software projects evolve and there are certain types of relationships between the actors. Later, Yu et al.\ \cite{Yu2008} adapted different symbiosis types, that might affect the evolution of projects, from biology and applied them to relationships between software projects. Furthermore, Scacchi and Alspaugh \cite{Scacchi2012} studied how different licenses affect on the ecosystem evolution. 

To summarize, while there are few studies addressing the evolution of software ecosystems, they represent different ends of the spectrum: On one corner, there are studies on the relationships and evolution of software code base; and on the other corner, studies have focused on visualizing interfirm relationships of companies with Social Network Analysis. 
To the best of the authors' knowledge, this study is unprecedented as it combines both of the existing schools of thought in the study of ecosystem evolution:  We study software ecosystems as a business network construction
 but acknowledge the importance of source code and address the evolution of the ecosystem and interfirm relationships through the developments in the shared codebase. In addition, we specifically focus on the co-evolution of competing software ecosystems. 
Generally, \emph{co-evolution} refers to cases where two entities affects to each others' evolution. Here, the entities are ecosystems and their actors.

\section{Empirical background}
\label{sec:empiricalbc}

The cloud computing business is dominated by a relatively small number of players, including 1) Amazon, a pioneer in cloud computing services selling the Amazon EC2; 2) Google, selling services around its Compute Engine (Google Compute);  and  3) Microsoft, heavily marketing cloud strategies based on its Azure cloud computing infrastructure (Microsoft Azure). The entrance costs for building and providing a public cloud computing infrastructure are very high as they often require global-distributed data-centers, fast and large accesses to the Internet backbone, much computing and storage power. Public cloud providers must provide very low latency -- after all, they are convincing its enterprise customer to move from  self-managed in-house computing infrastructures to vendor-managed computing infrastructures out there.

The leader of the cloud computing industry (i.e., Amazon, Google, and Microsoft) do not provide cloud infrastructure products, merely computing services. In practice and if there were no alternatives, all cloud computation 
would run in hardware and software infrastructures controlled by very few players. Such control from the cloud computing service provider locks-in its customers \cite{armbrust2010view}. 
Surprisingly, the leading product alternatives to Amazon EC2, Google Compute and Microsoft Azure are not commercial but rather four open-source projects. They include:  
1) OpenStack, our unit of analysis;  
2) CloudStack, backed by Citrix and the Apache Software Foundation;  
3) Eucalyptus, a system that is compatible with Amazon EC2 services and backed by many IT consulting firms;
and 4) OpenNebula, more present in the European markets and backed by C12G, a Spanish company. 
During our research, we perceived that many cloud computing vendors associated with the leading open-source cloud computing ecosystems to ease the pain of ``selling cloud computing services that are famous and infamous for their single-vendor locking mechanisms''. 

OpenStack is a software cloud computing infrastructure capable of handling big data. It is often offered as an IaaS (Infrastructure-as-a-Service) solution.  The development of this open-source software involves private companies (such as AT\&T, Canonical, Ericsson, IBM, Intel \emph{etc.}), public organizations (such as NASA, CERN, Johns Hopkins University \emph{etc.})  as well as independent, non-affiliated individuals.

We selected OpenStack as our case study subject due to four main factors. First, it is truly heterogeneous software ecosystem including start ups, high-tech corporate giants, non-profit and public organizations as well as individual software developers.  Second, it is highly inter-networked. That is, there are several companies and individual contributors working together. Thus, there is a rich data available for co-evolution. Third, its size is large enough for a meaningful study (more than 70.000 individual contributors and more than  600 supporting companies from 185 countries that have contributed with more than 20 million lines of source-code\footnote{See \url{https://www.openstack.org/community/}.}). Finally, it is well-studied \cite[see e.g.][]{teixeira2015lessons} and, therefore, there is a good amount of scholarly information published.

\clearpage

\section{Method}
\label{sec:method}

In this section, we present our research design. Given the multidisciplinary nature of our research approach which borrowed  significantly across disciplines, many interwoven methodological issues are disclosed. We employed a case study research strategy \cite{yin2011} that relied on naturally occurring data which emerged \emph{per se} on the Internet. Such data (e.g., web pages, wikis, blogs, public announcements, market-research reports, technical documentation, the software, source code repositories, videos broadcasted from the OpenStack summits, among many others data sources) are not a consequence of researchers' own actions, but rather are developed by the OpenStack community in their own pursuits of developing an open-source infrastructure for handling and storing big amount of data.  

Given the open-source nature of our focal unit of analysis, many but heterogeneous data regarding OpenStack is available. Therefore we have selected a novel approach by combining three well-known technique: mining software repositories (MSR) of OpenStack repository, Social Network Analysis (SNA) of the contributing developers, and qualitative analysis of archival data (QA). All within a mixed methods design, that reconstruct as well as visualize the evolution of the software ecosystem as a sequence of networks connectiong firms and individuals that jointly develop the OpenStack ecosystem. 

We started our efforts qualitatively  by searching publicly available data sources such as news articles, public announcements by companies, financial figures as well as press reports. Those helped us to create a picture of the cloud computing industry where OpenStack is a part of. In addition, we went through OpenStack documentation regarding how the software ecosystem is developed (i.e. the technical information) and  governed (e.g., structures,  policies, and procedures). While keeping in mind the limitations on the use of archival data~\cite{yin2011}, we gained valuable insights from OpenStack community and its surrounding industrial environment. After gaining an understanding of the surrounding industrial dynamics and understanding of how OpenStack software is developed, we extracted the developer and affiliation information from the publicly-available OpenStack Nova repository. Then, we created and analyzed the social network of the project by using the SNA guidelines given in~\cite{wasserman_social_1994}.

As in \cite{teixeira2015lessons}, we took advantage of naturally occurring digital trace data (i.e., the OpenStack Nova project repository and its \emph{changelog}) and built cooperative social networks that were analyzed using  a variety of tools: \emph{Gephi}, \emph{Visone}, and the \emph{sna} and \emph{statnet} statistical modules for \emph{R}. To better explore cooperation at the ecosystems level, we also modeled  cooperative relationships in the tri-dimensional (3D) space using \emph{Blender}. We mined evidence of cooperation from the source code and by visualized the social structures with SNA. This revealed the cooperation in the OpenStack ecosystem and we later enrichment this data with qualitative information from the public sources used in QA. The use of all these methods were helpful in terms that they both showed the social structures as well as helped to explain them.

We highlight the visualization of the collaboration network. The changes in this network, over time, show the dynamics among the OpenStack ecosystem. We aim to understand the visualized networks with the information gathered from the industry in previous steps. In this, we follow prior work \cite[e.g.][]{basole_visualization_2009,teixeira2015lessons} done in multi-disciplinary settings. 

\section{Results}
\label{sec:results}

We present our results in a chronological narrative format. The textual narrative is complemented with visualizations that capture the evolution of the OpenStack ecosystem.  Besides richly describing the evolution of the OpenStack ecosystem, we also attempt to interpret such evolution and explain it by employing multiple theoretical lenses. Our analysis aggregates both empirical and theoretical issues that are later addressed in the discussion section. 

We start with the words of, at that time Senior Vice President and General Manager of Rackspace, \emph{Jim Curry}. In this, the first public disclosure of the OpenStack project, Curry emphasizes the roles of NASA's and Rackspace's roles as initial contributors to the project -- that is, it is built with experienced partners and the project did not start from scratch.

\blockquote{
''Our mission statement says this:
   \emph{To produce the ubiquitous Open Source Cloud Computing platform that will meet the needs of public and private clouds regardless of size, by being simple to
    implement and massively scalable.} \\ 
That is a big ambition. The good news is that OpenStack is starting with code contributions from two organizations that know how to build and run massively scalable clouds -- Rackspace and NASA.'' 
--- Jim Curry, founder of OpenStack on behalf of Rackspace, 19 July 2010\footnote{See \url{https://www.openstack.org/blog/2010/07/introducing-openstack/}.}}

The footsteps of Rackspace in NASA started as a supplier of Anso Labs.  A startup company which was later acquired by Rackspace on February 9, 2011\footnote{See \url{https://gigaom.com/2012/05/24/nasa-backs-off-openstack-development/}.}. 
Before OpenStack, Anso Labs and Rackspace  have been working in Nebula -- a Federal cloud computing platform. Nebula emerged at NASA Ames Research Center at Moffett Field, California in 2008. It allowed NASA researchers to manage the computation of data-intensive research projects in a cloud computing way. The design of Nebula  reflected the growing popularity of the Amazon Web Services (AWS) cloud computing environments. 

\blockquote{ ``Nebula's architecture is designed from the ground up for interoperability with commercial cloud service providers such as Amazon Web Services, offering NASA researchers the  ability to easily port data sets and code to run on commercial clouds.''
--- NASA under the Open Government Initiative, 7 April 2010\footnote{See \url{https://www.nasa.gov/pdf/440932main_Nebula.pdf}.}}

The NASA Nebula team started by adopting the Eucalyptus open-source cloud computing infrastructure  (now a competitor of OpenStack), as it resembled the EC2 compute cloud and S3 storage cloud technologies from Amazon. However, NASA faced scalability issues. After all, NASA demands computing and storage were very high. Nebula could accommodate files as large as eight terabytes. Furthermore,  Nebula could support only an individual file system of 100 terabytes. As an example, the maximum for Amazon EC2 file size was just one terabyte and and for file system size was also one terabyte\footnote{See \url{https://www.nasa.gov/open/nebula.html}.}. 

In addition of scaling requirements to handle big data, NASA engineers were not happy with the `open-core' business model strategy of Eucalyptus Systems Inc to monetize its cloud computing software ecosystem. According to NASA, Eucalyptus-based clouds were not entirely open-source.

\blockquote{``NASA engineers attempted to contribute additional Eucalyptus code to improve its ability to scale, they were unable to do so because some of the platform's code is open and some isn't. Their attempted contributions conflicted with code that was only available in a partially closed version of platform maintained by Eucalyptus Systems Inc., the commercial outfit run by the project's founders.''
--- Chris Kemp, NASA chief technology officer, 20 July 2010\footnote{See \url{https://www.theregister.co.uk/2010/07/20/why_nasa_is_dropping_Eucalyptus_from_its_nebula_cloud/}.}.}

\begin{figure}[!htb]
    \centering
            \centering
        		\includegraphics[keepaspectratio=true,width=0.9\textwidth]{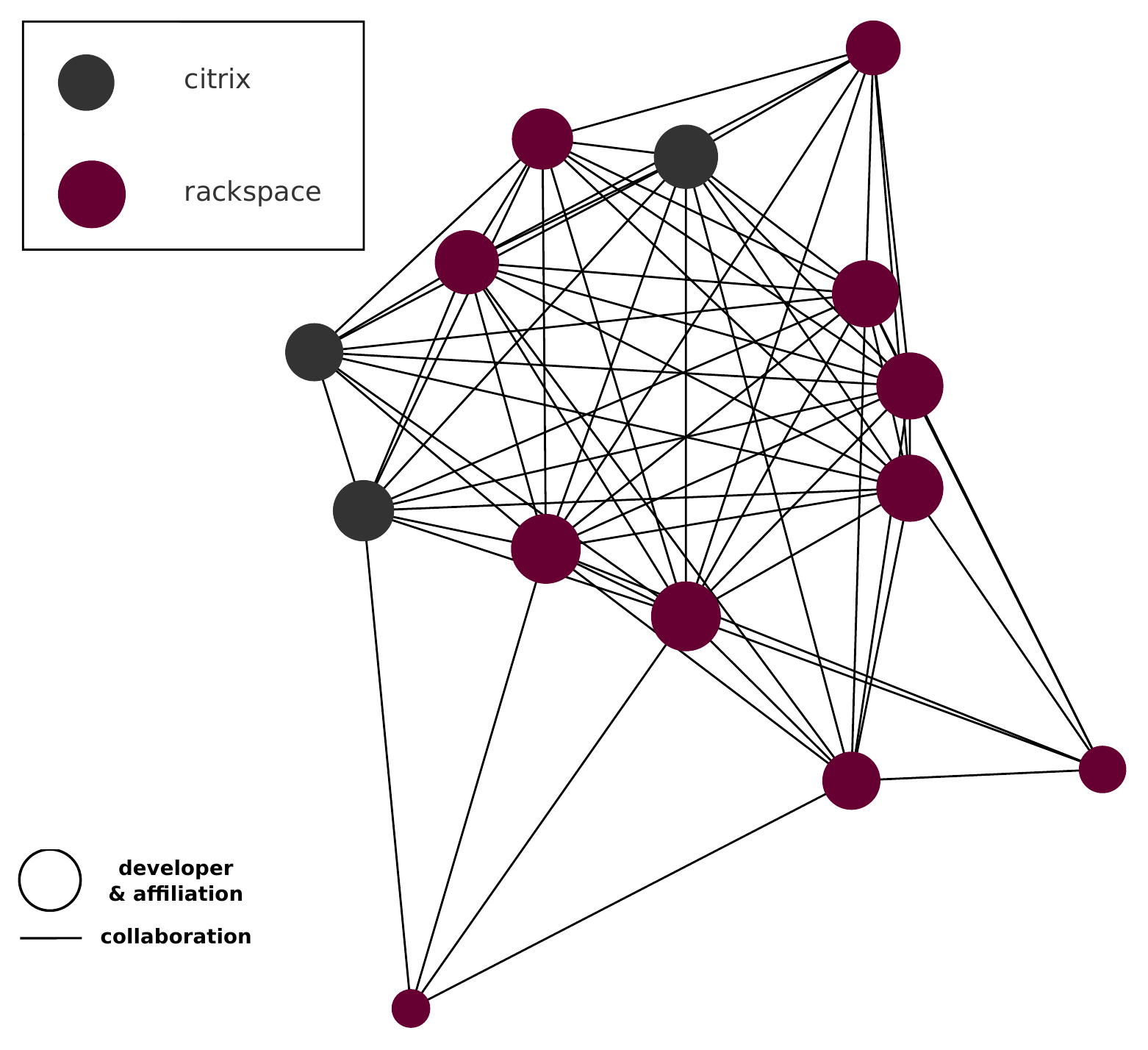}
        \caption{\texttt{Austin} $\rightarrow$ \texttt{Bexar}  \cite{teixeira2015lessons}.}
        \label{fig:Bexar}
\end{figure}

\begin{figure}[!htb]
            \centering
				\includegraphics[keepaspectratio=true,width=0.9\textwidth]{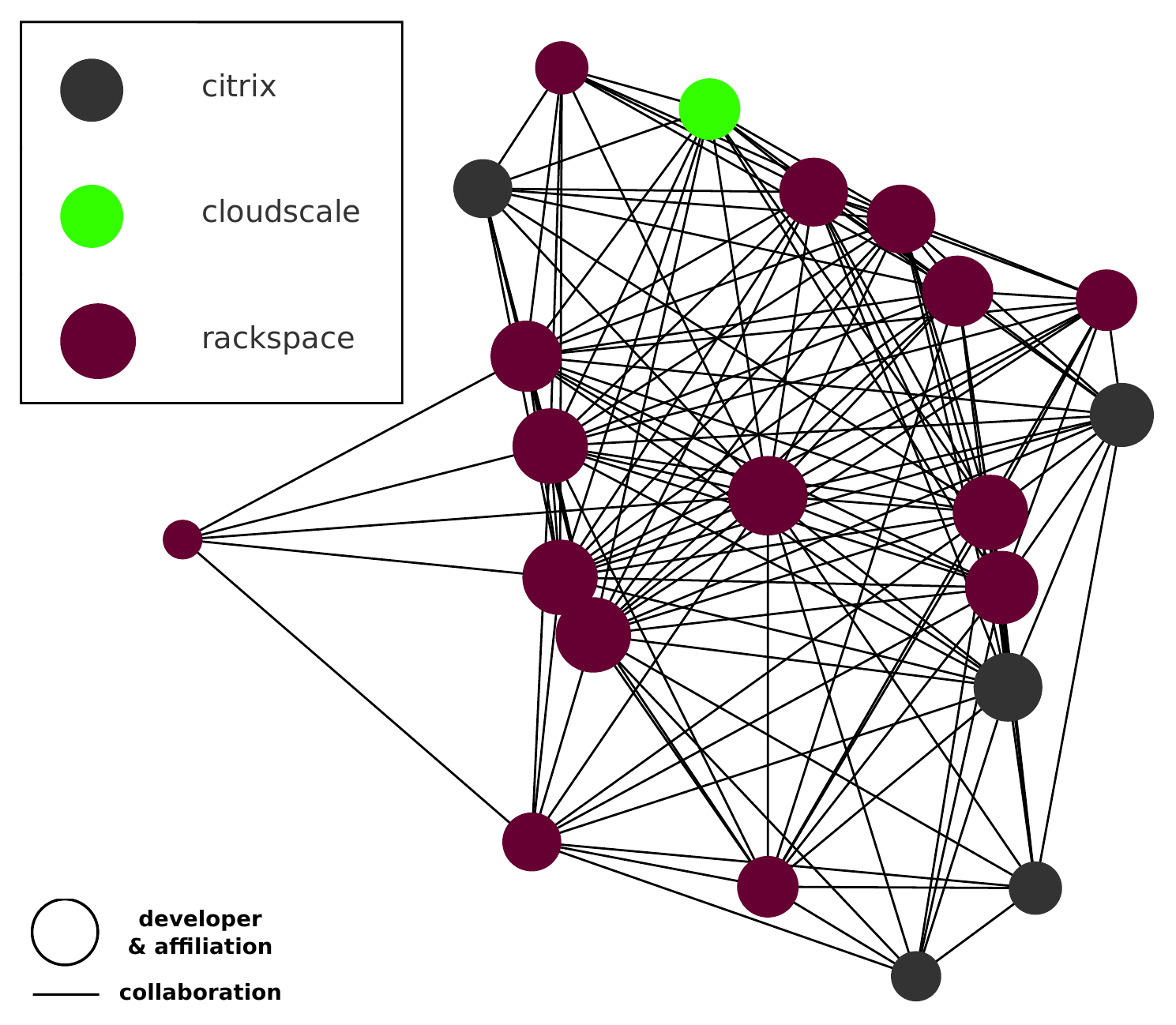}
		\caption{\texttt{Bexar} $\rightarrow$ \texttt{Cactus} \cite{teixeira2015lessons}.} 
		\label{fig:Cactus}
    \end{figure}

\begin{figure}[!htb]
    \centering
    	\centering
	  \includegraphics[keepaspectratio=true,width=0.9\textwidth]{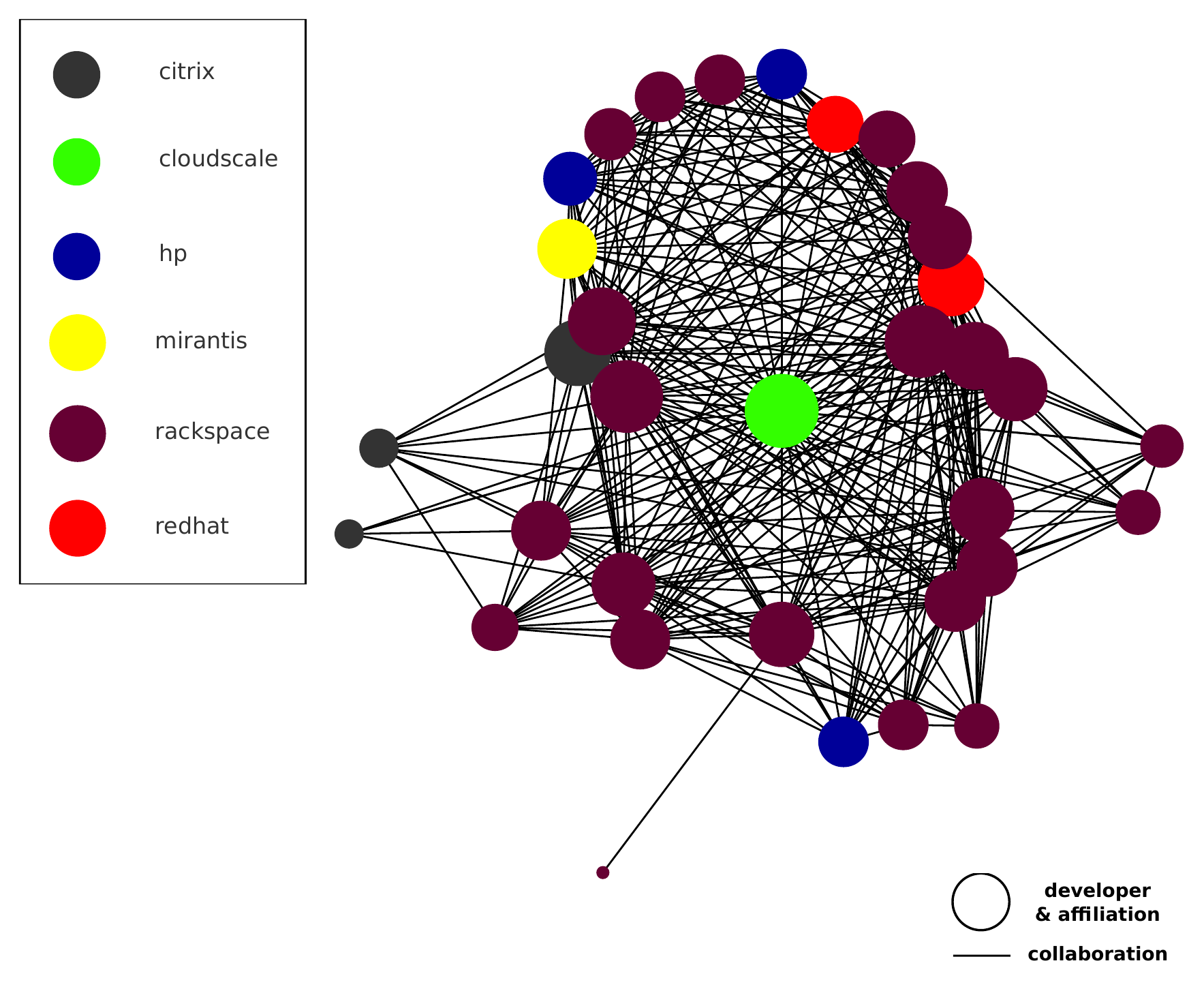}
		\caption{\texttt{Cactus} $\rightarrow$ \texttt{Diablo} \cite{teixeira2015lessons}.} 
	\label{fig:Diablo}
\end{figure}

As argued in prior related research \cite[see][]{teixeira2015lessons,teixeira_et_al_icis2016}, the visualizations in 
\Cref{fig:Bexar,fig:Cactus,fig:Diablo}\footnote{Please note that all figures are encoded as Scalable Vector Graphics, therefore readers can freely zoom in and zoom out for a better visualization of the networks.} helps us to understand how the cloud industry's actors cooperate in OpenStack.
Such visualizations, obtained with combining MSR and SNA,  helps us to visualize the evolution of the software ecosystem as an evolving complex network of companies and individuals interacting with each other to develop complex\footnote{Complex as it involves different programming languages, different operating systems, dozens of different hardware configurations, hundreds of firms,  thousands of software developers, and over one million of lines of code} software. The diameter of a node reflects its \emph{degree-centrality} -- in other words, a large node depicts a well-connected developer. The value of degree-centrality is a sum of the number of adjacent nodes with which a focus node is connected to. Thus, a high degree-centrality value, the more likely the developer is to be cooperating with other developers.


To start with,  \Cref{fig:Bexar} captures the cooperation in the OpenStack Nova project from the \texttt{Austin} (October 21\textsuperscript{st} 2010) to the \texttt{Bexar} (February 3\textsuperscript{rd} 2011) release.
The figure illustrates the cooperation between individual software engineers and their affiliated companies. For example, as shown by the figure, Citrix had three developers working on the project together with Rackspace.

Citrix's, who had worked before with Rackspace in Desktop visualization technologies\footnote{See \url{https://ir.rackspace.com/phoenix.zhtml?c=221673&p=irol-newsArticle&ID=1608440}.}, aim was to ensure that their XenServer platform would be included in OpenStack's future plans.  

\begin{quotation}
\noindent  ``As a longtime technology partner with Rackspace, Citrix will cooperate closely with the community to provide full support for the XenServer platform and our other cloud-enabling products.''
--- Peter Levine, SVP and GM, Citrix, 19 July 2010\footnote{See \url{https://www.rackspace.com/blog/newsarticles/rackspace-open-sources-cloud-platform}.}. 
\end{quotation}

Our second visualization, in~\Cref{fig:Cactus}, captures the cooperation from the \texttt{Bexar} (February 3\textsuperscript{rd} 2011) to the \texttt{Cactus} (April 15\textsuperscript{th} 2011) release. The figure illustrates the entrance of a new actor, a developer from the company Cloudscaling. 

The company started in 2006 with personnel previously working for Amazon and VMWare. It started by selling customized cloud infrastructures for large service providers. For example, Cloudscaling had Korea Telecom as an early customer. In 2010, the company shipped an OpenStack-based storage cloud to Korea Telecom. It was first OpenStack delivery without Rackspace.  Together with Mirantis, they were among the first pure-play OpenStack firms deploying OpenStack-based in-premise private clouds  (e.g. Korea Telecom and PayPal). While CloudScaling kept a strategy of compatibility with Amazon EC2 APIs, Mirantis was more on the position that OpenStack should not follow the designs of its competitor but challenge it\footnote{See  presentation entitled ``OpenStack Co-Opetition: A View from Within'' from Boris Renski (co-founder and chief marketing officer of Mirantis) presented on 04 Nov 2013  at the OpenStack summit, Hong Kong. Available on youtube at \url{https://www.youtube.com/watch?v=i7HXu2abNj0}.}. 

\begin{quotation}
\noindent ``We are  introducing a cloud infrastructure suite of products that essentially delivers an Amazon Web Services-like cloud, but on a customer's premise.''
--- Michael Grant, Cloudscaling's CEO, 9 February 2012\footnote{See Nancy Gohring  news article at \url{http://www.infoworld.com/article/2619192/}.}.
\end{quotation}

\begin{figure}[!htb]
 	\centering
	\includegraphics[keepaspectratio=true,width=0.9\textwidth]{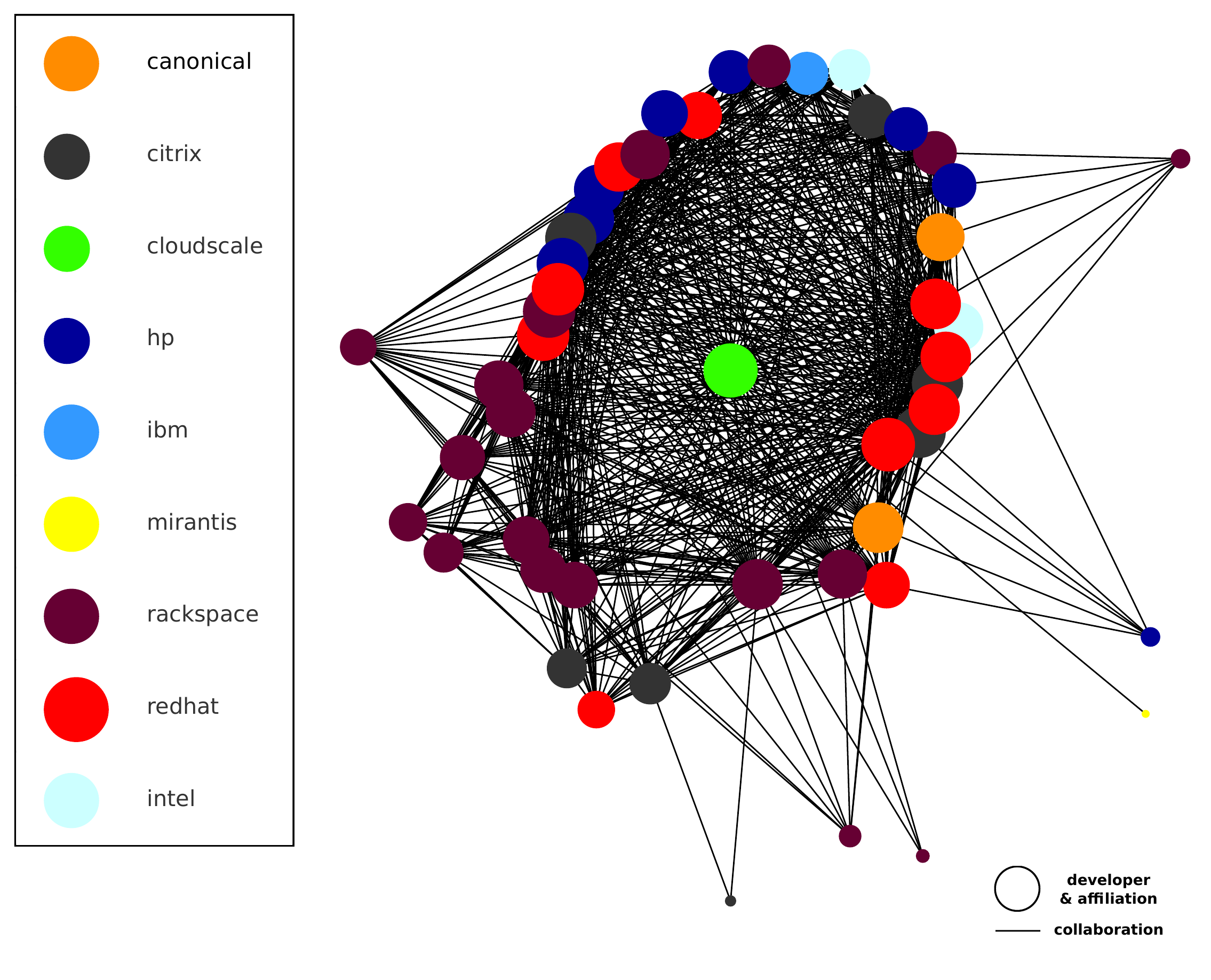}
		\caption{\texttt{Diablo} $\rightarrow$ \texttt{Essex} \cite{teixeira2015lessons}.} 
	\label{fig:Essex}
    \end{figure}

Our visualizations in~\Cref{fig:Diablo,fig:Essex} capture cooperation from the \texttt{Cactus} (April 15\textsuperscript{th} 2011 ) to the \texttt{Diablo} (September 22\textsuperscript{nd} 2011) release  and cooperation from the \texttt{Diablo} (September 22\textsuperscript{nd} 2011) to the \texttt{Essex} (April 5\textsuperscript{th} 2012) release. HP and IBM (large IT companies), Mirantis  (an OpenStack pure-play startup), Red Hat (a Linux operating system distribution's vendor), Canonical (company behind the Ubuntu Linux distribution), VMware (an expert on the virtualization software and services) and Intel (selling CPUs that powered cloud infrastructures) got involved in the coopetitive\footnote{Coopetitive as firms within OpenStack cooperate and compete simultaneously. See  \cite[p. 6]{teixeira2015lessons} for a relational map of competition among OpenStack firms.} software project. 

Mirantis, founded in 2011,  marketed itself as a ``pure-play'' OpenStack company. The startup started collaboration early with Red Hat.  Besides cooperating in the development of OpenStack, both firms partnered in implementation and integration services at common customers\footnote{See \url{https://www.redhat.com/en/about/press-releases/red-hat-and-mirantis-partner-across-products-and-services}.}. Mirantis was involved in the early deployments of OpenStack at large enterprises such as Paypal, AT\&T, Comcast, and Wells Fargo among others. 

In the meantime, HP launched an OpenStack-based cloud computing services. The company started marketing itself as the leading organization behind the project. In addition, they marketed OpenStack as free of single-vendor locking as there were a full ecosystem behind the project\footnote{See \url{https://www.openstack.org/foundation/companies/profile/hewlett-packard-enterprise}.}. At that time, it was the only cloud computing solution with a such promise. After all, cloud computing services are known by locking-in its customers \cite{armbrust2010view}.

With a very good track of contributions to open-source projects IBM (top contributor to the Eclipse IDE project), RedHat (top contributor to the Linux kernel) and Canonical (top contributor to the GNOME Linux Desktop project) also joined OpenStack. In common, all those companies had much expertise on Linux, the host operating system of OpenStack. RedHat and Canonical aimed at being the \emph{defacto} host operating system for OpenStack-based clouds\footnote{See \url{https://www.openstack.org/blog/2013/11/openstack-user-survey-october-2013/}.} 

IBM entered with force in OpenStack and showed much commitment to the platform. Besides contributing with much source-code to the project, it helped many of its customers to deploy openStack. Moreover, it entered into the public cloud business with OpenStack as well. In the case of IBM, as well with HP and Intel, money could be made by selling complementary hardware optimized for OpenStack. On the space of virtualization technologies, VMware did not want to lose ground to Citrix, and its contributions to OpenStack ensured compatibility with its vSphere,  NSX, vSOM and vCloud offering\footnote{See \url{http://www.vmware.com/products/openstack.html}.}. 

\section{Discussion}
\label{sec:disc}

In this section we discuss our most significant results.  The structure of the discussion reflects our mixed-methods analytical approach where we attempted to maker sense of the retrieved social network visualizations capturing cooperative relationships within a complex software ecosystem. After all, ``the fundamental quest of SNA is to understand the structure of the network'' \cite[p. 36]{carrington2014_review_of_sna}. 

\subsection{A theoretical and empirical evolutionary approach to software ecosystems}

In order to understand and explain why the retrieved social network visualizations took such topology and not other, much theoretical and empirical background knowledge was required. The use of certain theory to understand and explain our results was complemented with our understanding of the competitive cloud computing industry in which OpenStack is embedded, as well as with our understanding of how OpenStack is developed and governed. Besides literature directly adressing software ecosystems \cite[c.f.][among many others]{Jansen2009a}, our explanation integrated as well with theory on the embeddedness of business relationships \cite{uzzi1997social},  management of the portfolio of business relationships  \cite{hoffmann2005manage}, cooperation among competitors \cite{rochet2002cooperation},  materiality of technology \cite{Zammuto_et_al2007}, innovation and intelectual property  \cite{guildea2016app} and multi-homing strategy \cite{Landsman2011}.  Empirical and theoretical knowledge was complementary -- we could not explain the complex evolution of the OpenStack ecosystem without much knowledge on the surrounding industrial background of OpenStack or knowledge on the internal socio-technical practices by key actors within the development of OpenStack. Furthermore, a theory was fundamental to derive why the cooperative relationships (captured with SNA from the source-code) within OpenStack evolved in one way and not other. 

By tentatively explaining the retrieved networks, and by focusing on OpenStack in relation to other software ecosystems in the industry, we identified seven mechanisms that shaped the evolution of OpenStack. Such mechanisms are not internal to OpenStack,  but are rather enacted by other software ecosystems in the industry. In other words, we identified different ways on how do software ecosystems mutually co-evolve. We found seven mechanisms by which software ecosystems mutually influence each other --- but we do not reject the existence of others. The  mechanisms reported here can be seen as that drive the evolution of a software ecosystem in relation to others. Interesting enough, some of those identified causal mechanisms are enacted by competing software ecosystems. 

\subsection{Mechanisms of co-evolution among software ecosystems}

In the following, we will present the identified co-evolution mechanisms. The list is not complete and further work is needed to validate the mechanism. In addition, some of the mechanisms may overlap partially; however, we have decided to present them separately in order to validate or reject them in further studies.

\smallskip

\noindent $\alpha$)  \textbf{Sedimentation and embeddedness of business relationships:}
When analyzing the complex history of OpenStack, we quickly notice that prior business relationships had much impact on the evolution of OpenStack.  For example, Rackspace entered in the  NASA Nebula project (the precursor of OpenStack) as a supplier of Anso Labs (a company that it was later acquired by Rackspace). Another early contributor to OpenStack, Citrix, worked before with Rackspace in the development of Desktop visualization technologies before embracing OpenStack.  As pointed out by strategic management theory,  business partnerships often accumulate in a process of sedimentation \cite[c.f.][]{wassmer2010manage}. Moreover, actors tend to cooperate with actors that they had previously worked with and  tend to buy from existing suppliers (over  new suppliers) --- all in the so-called paradox of the  embeddedness of business relationships \cite[c.f.][]{uzzi1997social}. 

\smallskip

\noindent $\beta$) \textbf{Strategic management of the portfolio of business relationships:}
Many of the firms contributing to OpenStack (e.g., IBM, HP, and RedHat) manage a vast portfolio of business partnerships. However, the  value from a stand-alone partnership may not necessarily be value-creating from the overall portfolio perspective. Potential synergies between multiple alliances must be balanced to mitigate conflicts with other alliances \cite[c.f][]{hoffmann2005manage}. Companies such as IBM and HP needed to show commitment to  OpenStack as contributions to other cloud computing ecosystems could potentially damage cooperation. Firms that strategically engage with a software ecosystem might not be able to participate in competing others to not damage existing relationships. After all, the ``friends of my enemies are my enemies'' and the ``enemies of my enemies are my friends''. 
\smallskip

\noindent $\gamma$) \textbf{Firms values and reputation as a partner:}
Another found mechanism with significant impact on the co-evolution of software ecosystems is each firm values and reputation as a good partner. Something especially important in cooperation among competitors \cite[c.f.][]{rochet2002cooperation,teixeira_et_al_icis2016}.  In the case of OpenStack, RedHat had a good reputation as a top contributor to Linux;  IBM had a good reputation as a  top contributor to Eclipse;  Mirantis had a good reputation in deploying OpenStack to its customers while contributing back upstream to the community. On the other hand, Eucalyptus had lost some of its reputation of working in a truly open-source way by closing parts of the Eucalyptus cloud computing ecosystem. The evolution of software ecosystems is then a function of the values and reputation of existing and possible participants.  With the loss of reputation,  a player might disappear from an ecosystem while becoming unwelcome in others as well. 

\smallskip

\noindent $\delta$) \textbf{Core technological architecture}:
OpenStack is only functional with a ``host OS''.  At the early days of OpenStack,  Citrix and Rackspace welcomed much the expertise on Linux from RedHat and Canonical. In the cooperative side, they could optimize the ``host OS'' to better run OpenStack. On the other hand, on the competitive side, this two companies competed with others for customers with the sales argument that ``we know Linux, we know OpenStack and we are the only ones that can support both''. This reminded us that the stacking of architectural layers influences the software ecosystem evolution.  The materiality of a software ecosystem \cite[c.f.][]{Zammuto_et_al2007} influences the materiality of other software ecosystems. In the case of OpenStack, the architecture of the OpenStack core was very much influenced by the architecture of Linux, the architecture of Eucalyptus (now a competitor) and consequently the architecture of Amazon EC2.        

\smallskip

\noindent $\epsilon$) \textbf{Design of external APIs}: 
We also noted that not only the technological architecture of a software ecosystem influences its co-evolution with others, but also the design of its external APIs. At the early days of OpenStack, the design of the APIs from OpenStack, Eucalyptus, and Amazon EC2 converged. As the open-source cloud computing alternatives matured, Eucalyptus pursued compatibility with the established Amazon AWS APIs while OpenStack opted to diverge and provided interfaces to its computing and storage services in a distinct way. At that time, many decided to get away from OpenStack and move to Eucalyptus and CloudStack because its API was very different from the established  Amazon AWS APIs. Many customers wanted to easily move applications from Amazon EC2 to their own private clouds and the other way around -- Amazon AWS still remains the leader in public cloud services. 

\smallskip

\noindent $\zeta$)  \textbf{Competitive replication of new functionality}:
Competition forces players to copy functionality from each other to keep the pace. Besides the existence of many intellectual property protection mechanisms, this happens quite often in the software industry  \cite[c.f.][]{guildea2016app}. OpenStack started by implementing (and improving) many of the functionalities provided by Eucalyptus and Amazon E2C. Moreover, whenever firms tried to embed OpenStack within a proprietary product (e.g., a new proprietary cloud orchestrator that embedded OpenStack in it), OpenStack implemented a ``new, official and open-source version'' of it (i.e., OpenStack launched a new sub-project implementing the official orchestrator for OpenStack). In other words, OpenStack expanded its core by replicating complements. After all, copying an idea and making it open-source is often a more powerful tactic than copying open-source software and making it proprietary. 

\smallskip

\noindent $\eta$) \textbf{Complementors' multi-homing:}  
Due to the nature of software, the complements of ecosystems are often intangible. This means that these products and services (that add much value to the overall ecosystem) can be transferred or moved to a different ecosystem and setting with a relatively little effort. The phenomenon where a single complementing actor is offering his or her products or services to two or more ecosystems at the same time is called  \emph{multi-homing} \cite[c.f.][]{Rochet2003,Hyrynsalmi2012c,Hyrynsalmi2016d}; the opposite strategy is known as \emph{single-homing}. For example, due to the success of the `Flappy Bird' mobile game, it was quickly copied into all major mobile application ecosystems, and beyond \cite{Hyrynsalmi2014c}. Within OpenStack,  Xen from Citrix and ESXi from VMWare are Hypervisors\footnote{A hypervisor is either a software or a hardware solution that creates, follows and runs virtual machine instances.} that work both in the OpenStack and Eucalyptus software ecosystems. The contributions that shape the ecosystem evolution, are often due to the complementors interest in making their offering available across different ecosystems. Citrix and VMWare wanted to be sure that their hypervisors run on different cloud computing platforms. In the realm of CPUs, Intel and AMD, also contributors to the OpenStack, are also interested in making sure that their CPUs work across different cloud computing platforms --- therefore they contribute to most open-source cloud computing software ecosystems. The same happens with Cisco, Juniper Networks, IBM and HP among other vendors of networking technology. 

\subsection{Implications and limitations of the study}

Besides richly narrating the evolution of a software ecosystem, our  focal perspective of attempting to understand the evolution of a software ecosystem as a function of other software ecosystem extends the literate on software ecosystems evolution \cite{basole_visualization_2009,Hanssen2012b, Yu2008}. 

Future research towards a deeper understanding of ecosystems' evolution should  acknowledge that ecosystems do not evolve in insulation. Careful analysis of a software ecosystem evolution should take in consideration other software ecosystems as well, including competing ones. We argue then that, in order to understand the evolution of an ecosystem, we need to look way beyond it. New methodologies, capable of capturing inter-ecosystem dependencies, are needed to addresses such findings \cite[c.f.][for recent advancements in this direction]{basole2015understanding,teixeira_et_al_icis2016}.

Naturally, there are certain limitations for this study. First, this study uses a single case study research design to identify mechanisms of co-evolution. Therefore, it is likely our list of mechanisms is not full and further work is needed in order to validate the identified mechanism as well as to find new ones. In addition, we used one ecosystem as a focal point and studied co-evolution from its point-of-view. While we selected the case study ecosystem carefully, there are threats involved in the single case study research design. In future work, multiple ecosystem point-of-view should be used to validate our results.

Second, we selected an open-source software ecosystem as the case subject and generalizing the results to the other kinds of ecosystems should be done with care. There are some previous discussion on the limits on generalizing results from different kinds of software ecosystems to another kinds \cite[c.f.][]{Hyrynsalmi2014c,Suominen2016a}; however, the results of this study does not heavily rely on a certain ecosystem type. Therefore, they should be generalizable to, at least, open-source software ecosystem and, with some limitations, to general type of software ecosystems.

Third, we used developers' point-of-view in studying co-evolution of software ecosystems. Another option could be study the business connections between the participating companies \cite[c.f.][]{Basole2011,basole2015understanding}. However, as software ecosystem are built from the developers' point-of-view \cite{Manikas2016a}, our decision seem to justifiable. Nevertheless, the further studies should pay attention also on other perspectives of ecosystems' co-evolution.

\section{Conclusion}
\label{sec:con}

 Our findings contribute to a deeper understating of the evolution of software ecosystem. We found that a software ecosystem co-evolve with other software ecosystems in at least seven different ways.  Understanding the evolution of a software ecosystem requires an especially astute awareness of its competitive landscape as well as knowledge on its internal socio-technical practices. Research addressing the evolution of software ecosystem should, therefore, acknowledge that software ecosystems entangle with other software ecosystems in multiple ways, even with competing ones. 

\section*{Acknowledgements}

Besides the three anonymous reviewers for ICSOB 2018, we would like to thanks participants from the 1st European Conference on Social Networks (INSNA EUSN 2014) \cite{Teixeira2014eusn}, the 10th International Symposium on Open Collaboration (ACM OpenSym 2014) \cite{openSym2014},  the 2nd IEEE International Conference Big Data (IEEE Big Data2014) \cite{Teixeira2014bigdata}, the 38th Information Systems Research Conference in Scandinavia (IRIS 2015)  \cite{IRIS2015}, and the 37th International Conference on Information Systems (AIS ICIS 2016) \cite{teixeira_et_al_icis2016} for early feedback on our research efforts.

This research was partially funded by the Fundação para a Ciência e a Tecnologia (grant SFRHBD615612009), Liikesivistysrahasto (grants 5-3076 and 8-4499), Marcus Wallenberg Säätiö (grant ``open-coopetition R\&D management strategy'') and Academy of Finland (decision no. 295743). We thank also the University of Turku, the Tampere University of Technology and Åbo Akademi for providing much of the needed research infrastructure.
Finally, we would like to thank the R, Gource, Gephi, Tulip, Blender, and the OpenStack open-source communities for developing cool and research-friendly software.

\begin{backmatter}

\section*{Competing interests}
  The authors declare that they have no competing interests.

\section*{Author's contributions}
    JAT conducted the data collection. SH participated in the analysis and contributed equally to the crafting and writing of this article.

\linespread{1.1}

\iftoggle{bibtex_natbib}{
\bibliographystyle{bmc-mathphys} \bibliography{references,hyrynsalmi,ecosystem}
}{}

\iftoggle{biber_biblatex}{
  \printbibliography
}{}

\end{backmatter}

\end{document}